\documentclass[12pt]{article}
\usepackage[top=1.25in, bottom=1.25in, left=1.25in, right=1.25in]{geometry}
\usepackage{lineno}
\usepackage[english]{babel}
\usepackage{placeins}
\usepackage{amssymb}
\usepackage{latexsym}
\usepackage{amsmath}
\usepackage{amsthm}
\usepackage{natbib}
\usepackage{graphicx}
\usepackage{setspace}
\usepackage{color}
\numberwithin{equation}{section}

\renewcommand{\vec}[1]{\mathbf{#1}}
\begin{document}
\title{Affine Reduction of Dimensionality: An Origin-Centric Perspective}
\date{}
\author{Robert L. Obenchain, Risk Benefit Statistics LLC, wizbob@att.net}
\maketitle
\begin{abstract}
We consider statistical methods for reduction of multivariate dimensionality that have invariance and/or commutativity properties under the affine group of transformations (origin translations plus linear combinations of coordinates along initial axes). The methods discussed here differ from traditional principal component and coordinate approaches in that they are origin-centric.  Because all Cartesian coordinates of the origin are zero, it is the unique fixed point for subsequent linear transformations of point scatters. Whenever visualizations allow shifting between and/or combining of Cartesian and polar coordinate representations, as in Biplots, the location of this origin is critical.  Specifically, origin-centric visualizations enhance the psychology of graphical perception by yielding scatters that can be interpreted as Dyson swarms.  The key factor is typically the analyst's choice of origin via an initial ``centering'' translation; this choice determines whether the recovered scatter will have either no points depicted as being near the origin or else one (or more) points exactly coincident with this origin.

\end{abstract}
\tableofcontents
\section{Introduction}
\label{sec: Introduction}
We discuss an intuitive alternative to classical Principal Components analysis that selects the (singular) affine transformation of the original scatter, visualized as N points embedded within a p-dimensional Euclidean space, that reduces its dimensionality to q $<$ p while attempting to reproduce all pair-wise, squared Mahalanobis inter-point distances.  We show that this criterion is not coordinate-free; it places extra emphasis upon reproducing the squared Euclidean distances of all N points from the origin.  Our origin-centric perspective on affine reduction of dimensionality encourages interpretation of the recovered scatter using either polar or Cartesian coordinates.

In traditional principal component and coordinate approaches to reduction of dimensionality, the origin is conventionally placed at the geometric centroid of the overall scatter by subtracting the observed arithmetic mean value from each given coordinate.  This convention can, optionally, be retained in the origin-centric approach discussed here.  On the other hand, the impact of details such as how the given coordinates were originally scaled and intercorrelated are minimized in this new, alternative approach.  This is accomplished by placing the given data scatter in its affine invariant canonical form, Obenchain(1971, 1972a).  Interestingly, this substitution makes reduction of dimensionality using conventional principal components analysis impossible; all one-dimensional orthogonal projections of the revised data have the same variance, and any two mutually orthogonal projections are uncorrelated.  In particular, this means that the ultimate impact of researcher initiatives, Glaeser(2006), in choice of included variables and of their initial scaling upon potential recovered configurations of low dimensionality is thereby greatly reduced.

The format of the paper is as follows. In the next three Sections, we introduce basic notation and define terminology related to Euclidean distance calculations, matrix norms in linear algebra, and origin-centric coordinate concepts.  Sections \ref{sec: example1} and \ref{sec: example2}, then display two relatively simple numerical examples.  We end this overview paper with some general discussion points in Section \ref{sec: Discussion}. The final four Sections, \ref{sec: affine invarcomm} to \ref{sec: plarge}, are somewhat more technical Appendices covering historical background information and some relatively advanced observations and conjectures.  

\section{Origin-Centric Coordinates and Euclidean Inter-point Distances}
\label{sec: origin-centric}
When a N $\times$ p matrix of finite real numbers, $\vec X$, is used as a Cartesian coordinate representation of N points in Euclidean space relative to p = 2 or more mutually orthogonal axes, the conventional view is that the location of the origin is not important.  After all, the Pythagorean theorem dictates that the squared distance between the i-th and j-th points in the scatter is given by
\begin{align}
\label{eqn: sqdist}
\vec d^2_{ij} &= (\vec x'_{i} - \vec x'_{j})(\vec x_{i} - \vec x_{j})\nonumber\\
    &= \vec x'_{i}\vec x_{i} + \vec x'_{j}\vec x_{j} - 2\vec x'_{i}\vec x_{j},
\end{align}
where $\vec x'_{i}$ denotes the i-th row of $\vec X$.  In the first line of the above expression, note that only differences in coordinates between points are involved.  It follows that  $\vec d^2_{ij}$ is, by definition, invariant under shifts in the location of the origin resulting from simple (additive) translations of coordinates.

The second line of equation (\ref{eqn: sqdist}) shows an equivalent expression containing origin-centric terms.  The first two terms are the squared Euclidean distances of the i-th and j-th points from their current origin, $\vec 0$.

Similarly, the full N $\times$ N matrix of inter-point squared distances can be written as
\begin{align}
\label{eqn: sqdistmtx}
\vec D^{(2)} &= ((d^2_{ij})) \mbox{  ...for all i and j from 1 to N} \nonumber\\
    &= (\vec d^2_0)\vec 1' + \vec 1(\vec d^2_0)' - 2\vec {XX'},
\end{align}
where $\vec d^2_0$ is the N $\times$ 1 column vector that consists of the ordered elements of $Diag(\vec {XX'})$.  In other words, the i-th element of $\vec d^2_0$ is the $\vec x'_{i}\vec x_{i}$ term of equation (\ref{eqn: sqdist}).  Thus $\vec d^2_0$ is the vector of squared Euclidean distances of all N points from their current origin, $\vec 0$.  The $\vec D^{(2)}$ matrix is symmetric, and its N diagonal elements are null.

Suppose now that a given $\vec X$ matrix is to be optimally approximated by a N $\times$ q matrix of finite real numbers, $\vec Z$, where q $<$ p but q is at least 1.  Specifically, suppose that the objective is to minimize the squared Frobenius norm of the difference between the $\vec D^{(2)}$ matrices computed from the $\vec X$ and $\vec Z$ coordinates:
\begin{align}
\label{eqn: sqdifnorm}
\vec {Norm}^2 &= \sum_{i=1}^N \sum_{j=1}^N (d^2_{ij}(\vec X) - d^2_{ij}(\vec Z))^2 \nonumber\\
    &= || {\rho}\vec 1' + \vec 1 {\rho}' - 2(\vec {XX'} - \vec {ZZ'})||^2 \nonumber\\
    &= 4||\vec {XX'} - \vec {ZZ'}||^2 + 2({\rho}'\vec 1)^2 + 2N{\rho}'\rho,
\end{align}
where $\rho = d^2_0 (\vec X) - d^2_0 (\vec Z)$ is the column vector of discrepancies in squared difference of points from the origin in $\vec X$ and $\vec Z$ coordinates.

In the special case of primary interest here, the $\vec {XX'}$ outer-products matrix ($N \times N$) is what is know as an Association matrix, $\vec A$, for ``individuals'' (points), Gower(1966), while the $\vec {X'X}$ inner-products matrix ($p \times p$) contains the corresponding Adjusted Sums-of-Squares and Cross-Products for ``variables.''  The objective in Principal Component Analysis (PCA) is to minimize $||\vec A - \vec {ZZ'}||^2$ where the $\vec X$ and $\vec Z$ matrices have each been ``centered'' as outlined below in Section \ref{sec: affinermd}.  This particular point is more fully discussed in an Appendix; see Section \ref{sec: center}.

In any case, minimizing $\vec {Norm}^2$ of equation (\ref{eqn: sqdifnorm}) is potentially quite different from ordinary PCA.  Minimizing $\vec {Norm}^2$ places emphasis upon balancing two separate criteria: [1] reducing the size of the $\rho$ vector, which contains the discrepancies in {\em squared distance from the origin} for all N points within the original $\vec X$ and derived $\vec Z$ configurations, and [2] reducing the PCA norm of the difference in outer-product matrices.  Tension between these competing objectives make choice of $\vec Z$ so as to minimize $\vec {Norm}^2$ of equation (\ref{eqn: sqdifnorm}) an origin-centric approach.

\section{Affine Reduction of Multivariate Dimensionality}
\label{sec: affinermd}

The {\em Singular Value Decomposition} of a N $\times$ p matrix consisting of the ``centered,'' real valued $\vec X$ coordinates of N points in a Euclidean space of dimension (linear rank) r can be written as:
\begin{equation}
( \vec I - \vec {1\gamma'} ) \vec X = \vec H \vec \Lambda^{1/2} \vec {G'}, \label{eqn: svdcenter}
\end{equation}
where the 1 $\times$ N row vector $\vec {\gamma'}$ is any generalized inverse of the N $\times$ 1 column vector of all ones, $\vec 1$, the semi-orthogonal matrix $\vec H$ is the N $\times$ r matrix of standardized {\em Principal Coordinates}, $\vec \Lambda^{1/2}$ is a r $\times$ r diagonal matrix of strictly positive singular values, and $\vec {G'}$ is a r $\times$ p semi-orthogonal matrix of component {\em Direction Cosines}.  When r = p, the $\vec G$ matrix becomes orthogonal (and invertible); $\vec {GG'} = \vec I$ of order p.

Every choice for $\vec \gamma$ is such that $\vec {\gamma'1} = 1$; it follows from (\ref{eqn: svdcenter}) that the semi-orthogonal matrix $\vec H$ is orthogonal to $\vec \gamma$.  This $\vec {\gamma'H} = 0$ restriction assures that r $\leq min(p, N-1)$.  More complete information about this critical choice of initial translation is given in the Appendix of Section \ref{sec: center}; choice of $\vec \gamma$ determines the location of the origin, $\vec 0$, of Cartesian coordinates.

Affine reductions in dimensionality are those in which the recovered $\vec Z$ configuration is of the restricted form: 
\begin{equation}
\vec Z = \vec H \vec B, \label{eqn: rqaffrmd}
\end{equation}
where $\vec H$ is the N $\times$ r matrix of (\ref{eqn: svdcenter}) and $\vec B$ is a r $\times$ q matrix of $rank(\vec B) = q < r$ consisting of real scalars.  In particular, any row of $\vec H$ that is null assures that the corresponding row of $\vec Z$ will also be null ...corresponding to a point coincident with the origin, $\vec 0$.

One of two forms of preliminary data ``standardization'' are typically applied prior to performing traditional PCA, but neither is nearly as ``drastic'' as replacing one's initial $\vec X$ matrix by its Mahalanobis principal coordinates matrix, $\vec H$, as in (\ref{eqn: rqaffrmd}).  Traditional ``mean centering'' uses the $\vec {1^+}$ vector, with each of its N elements equal to 1/N, as choice of $\vec {\gamma'}$ in (\ref{eqn: svdcenter}).  Traditional standardization to ``correlation form'' performs an addition step in which each column of the ``mean centered'' $\vec X$ matrix is then divided by its sample standard deviation.  This is essentially equivalent to the standardization in which the N elements in each column of $\vec X$ are made to not only sum to zero but also to have sum-of-squares set equal to 1; on the other hand, this standardization still allows the different columns of $\vec X$ to be (linearly) correlated or confounded.

In sharp contrast, the semi-orthogonal $\vec H$ matrix of (\ref{eqn: svdcenter}) is essentially r-dimensional from the perspective of traditional PCA.  Each column of this $\vec H$ matrix is orthogonal to the current choice of $\vec \gamma$ (defining a mean or multivariate median measure of location), has sum-of-squares set equal to 1, and is exactly uncorrelated with the derived coordinates along any direction strictly orthogonal to it.

For example, consider the 1-dimensional case, q = 1 in (\ref{eqn: rqaffrmd}), where the $\vec Z$ and $\vec B$ matrices become single column vectors, $\vec z$ and $\vec b$.  It follows that $\vec {\gamma'z} = 0$ while the corresponding sum-of-squares is $\vec {z'z} = \vec {b'H'Hb} = \vec {b'b}$, which is determined solely by the squared length of the $\vec b$ vector.  Similarly, any two different choices for this $\vec b$ vector yield uncorrelated $\vec z$ representations if and only if these two choices are mutually orthogonal vectors, $\vec {b_1'b_2} = 0$.

In other words, since PCA cannot be used to reduce the dimensionality of $\vec H$, one's primary hope may well be to adopt an origin-centric perspective based upon minimizing the squared norm of equation (\ref{eqn: sqdifnorm}). 

\section{Adding a Point at the Origin}
\label{sec: added origin}

If none of the rows of the $\vec H$ matrix of (\ref{eqn: svdcenter}) is null, then the given scatter contains no points coincident with the origin, as determined by the current choice of $\vec \gamma$ vector.  On the other hand, there will always be at least one point at the current origin whenever exactly one element of the current $\vec \gamma$ vector is a 1 and all other elements are null.

Any null row of the $\vec H$ matrix of (\ref{eqn: svdcenter}) can be relocated to its first row simply by renumbering the N points in the initial scatter.  The resulting $d^2_0 (\vec X)$ vector of (\ref{eqn: sqdifnorm}) that contains the squared Euclidean distances of all N point from their current origin will then naturally have a zero in its first row.  In fact, this $d^2_0 (\vec X)$ vector will also then be the first column of the $\vec D^{(2)}$ matrix of (\ref{eqn: sqdistmtx}), while its transpose will be the first row of this same matrix.

When a $\vec H$ matrix from (\ref{eqn: svdcenter}) contains no null row, a null row can be added as, say, row N+1. This augmented $\vec H$ matrix is still semi-orthogonal, and $\vec {H'H}$ is still the r $\times$ r identify matrix.  The corresponding augmented $\vec \gamma$ column vector ($(N+1) \times 1$) can be taken to have a zero as its final element, and the augmented $\vec D^{(2)}$ matrix of (\ref{eqn: sqdistmtx}) will have the original $d^2_0 (\vec X)$ vector as the above the diagonal elements in its last column ...and its transpose in the bottom row.

In summary, the $d^2_0 (\vec X)$ vector that contains the squared Euclidean distances of all N original points from their current origin is either already contained within the original N $\times$ N squared distances matrix, $\vec D^{(2)}$ of (\ref{eqn: sqdistmtx}), or else it will appear within the corresponding (N+1) $\times$ (N+1) augmented version of this matrix when a new point coincident with the origin is included.

In other words, the $d^2_0 (\vec X)$ vector is either an explicit or implicit part of every N $\times$ N matrix of squared distances, $\vec D^{(2)}$ of (\ref{eqn: sqdistmtx}).  In sharp contrast with PCA, efforts to minimize $\vec {Norm}^2$ of (\ref{eqn: sqdifnorm}) place special emphasis upon accurate reproduction of $d^2_0 (\vec X)$ values within the recovered Z-configuration.

\section{First Numerical Example: Six Points on a Plane}
\label{sec: example1}

Perhaps the most simple sort of ``interesting'' example for comparing PCA with our origin-centric approach is depicted as H coordinates (p=2 and N=6) in Figure 1.  When reducing this configuration to a single dimension (q=1), a singular affine transformation can produce either ``a pair of triplet points" or else a ``triplet of pairs.''  In fact, these particular configurations happen to correspond to two of the three local minima of $\vec {Norm}^2$ of equation (\ref{eqn: sqdifnorm}); see Figure 2.

\begin{figure}[htb]
\center{\includegraphics[width=\textwidth]{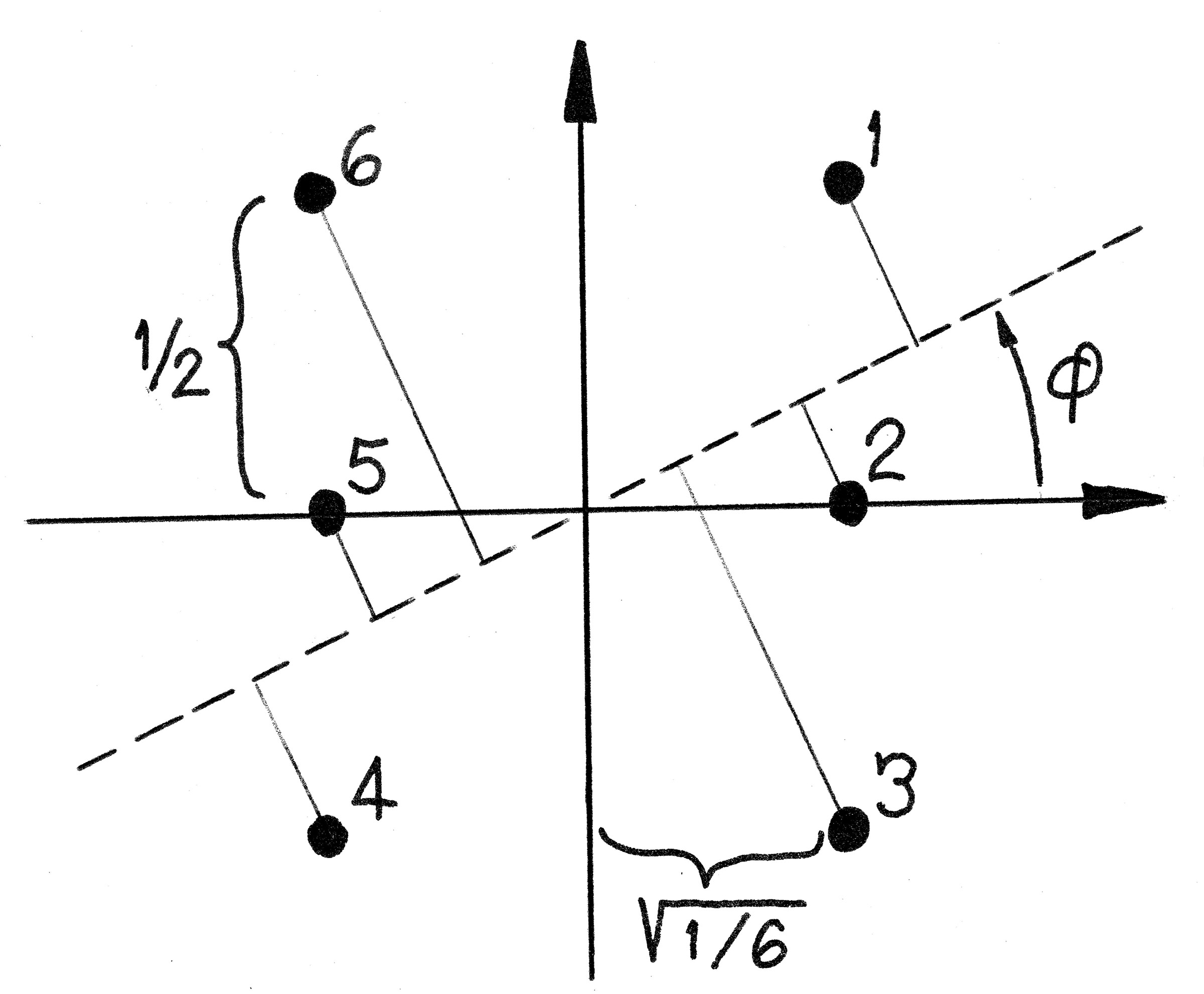}}
\caption{\label{fig:H6in2} Mahalanobis H Configuration for 6 Points in 2 Dimensions.  Numerical searches for the best q=1 representation can use, as their starting points, orthogonal projections of the 6 points onto the line through the origin making angle $\phi$ with the horizontal axis.  Initial values for the $\vec B$ matrix of (\ref{eqn: rqaffrmd}) are then simply 2-element column vectors of the form $(sin(\phi), cos(\phi))'$, and the objective of the search is to find the best length for $\vec B$ in each given direction.}
\end{figure}

This example is also interesting in the sense that, while the initial H-configuration contains no point(s) at the origin, the optimal q=1 dimensional recovered configuration [case (c) in Figure 2] contains 2 points coincident with the origin.

\begin{figure}[htb]
\center{\includegraphics[width=\textwidth]{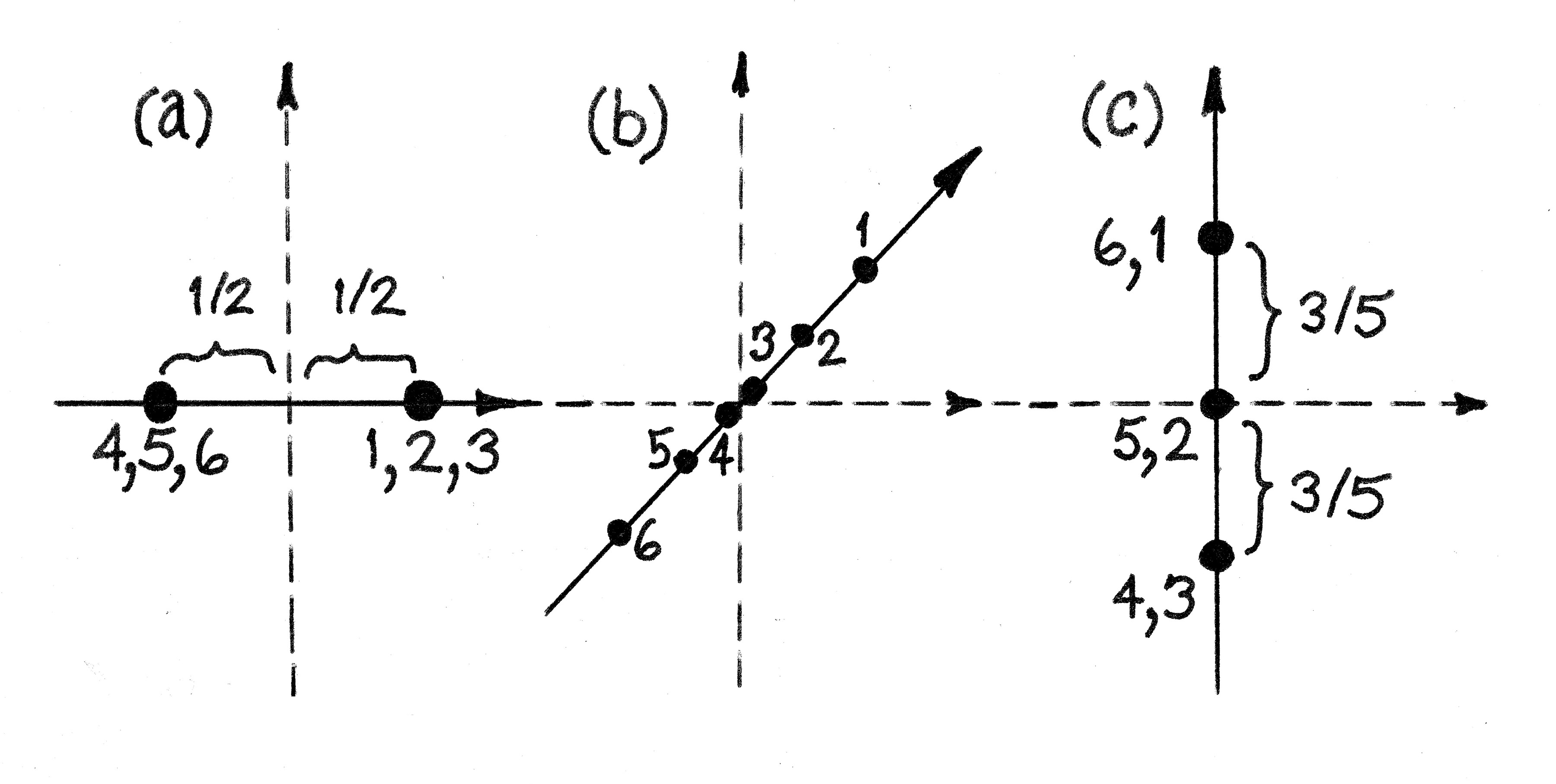}}
\caption{\label{fig:z6in2} Depiction of three Local Minima of $\vec {Norm}^2$ for Dimension q=1.  While a pair of local minima occur in case (a) at 3.50 and in case (b) at 4.80, the global minimum of 3.11 corresponds to case (c).}
\end{figure}

\section{Second Numerical Example: The Longley Data in q=2 Dimensions}
\label{sec: example2}

Here we use the infamous Longley(1972) dataset (p=6, N=16) to illustrate some key distinctions between the PCA and affine, origin-centric methods of (\ref{eqn: rqaffrmd}) for reduction of multivariate dimensionality. While Longley used seven variables in his ill-conditioned regression model, here we use YEAR = 1947 to 1962 only as labels for points, rather than as an exact ``linear trend" variable. The six US economic variables used here are $X_1 =$ GNP deflator, $X_2 =$ GNP, $X_3 =$ number Unemployed, $X_4 =$ size of Armed Forces, $X_5 =$ total Population and $X_6 =$ total Employed.   

\begin{figure}[htb]
\center{\includegraphics[width=\textwidth]{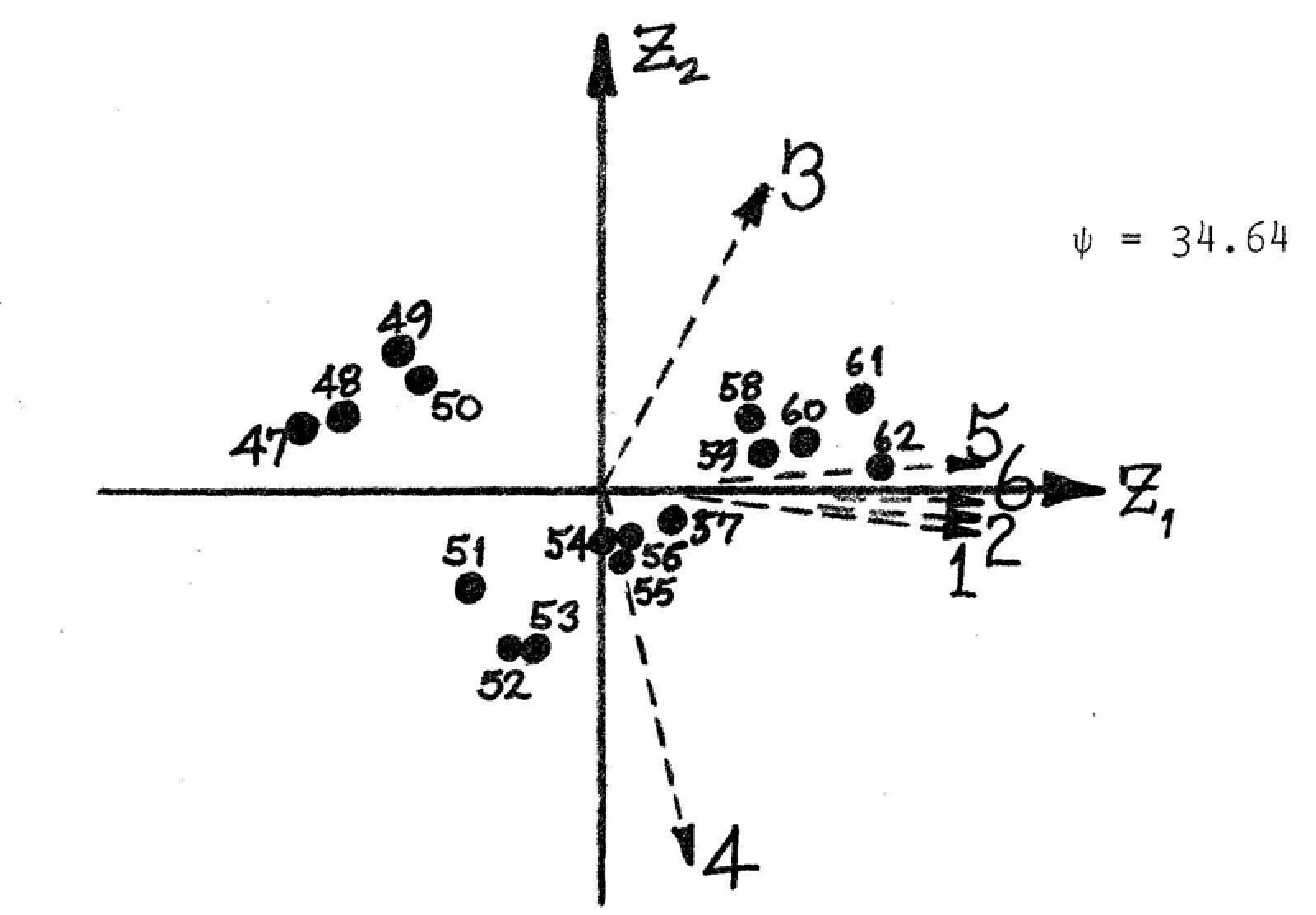}}
\caption{\label{fig:longPCA} The PCA solution of Dimension q=2 for the Longley data when standardized to ``correlation'' form.  It is clear from the labels on the points as well as from the direction cosines for Axes 1, 2, 5 and 6 that this recovered configuration is best interpreted as representing ``linear growth" over time.}
\end{figure}

\begin{figure}[htb]
\center{\includegraphics[width=\textwidth]{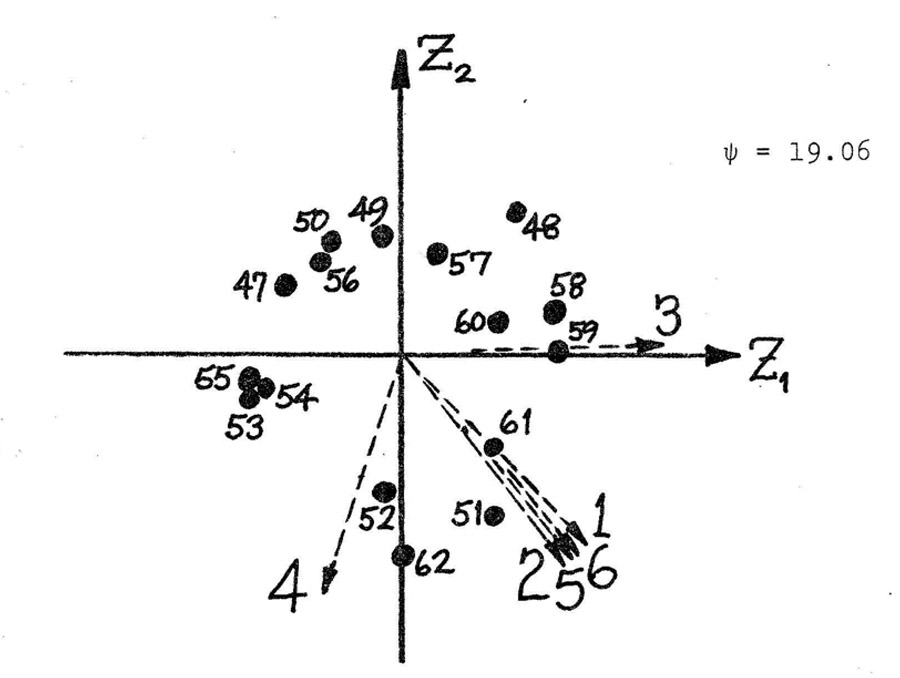}}
\caption{\label{fig:longAROD} This is the optimal origin-centric representation in q=2 dimensions for the Longley data when standardized to Mahalanobis H form.  The ``year'' labels on points suggest that time variation is being depicted here as clockwise or counter-clockwise rotations.  In particular, the direction cosines for Axes 3 and 4 suggest that these two measures are the true, primary determinants of year-to-year variation in q=2 dimensions.}
\end{figure}

This example is rather typical of cases I have explored, especially when N greatly exceeds p.  Specifically, the optimal affine reduction in Mahalanobis dimensionality depicted in Figure 4 has a ``Dyson swarm'' interpretation, featuring a characteristic ``hole" surrounding the explicit origin. 

\section{Discussion} 
\label{sec: Discussion}
In the terminology of Glaeser(2006), ``Researcher Initiatives'' can deliberately bias findings towards the perspective of one particular subset of stakeholders.  For example, it is well know that including several different quantitative X-variables that are surrogate measures of a single construct in a PCA tends to orient the first principal axis to represent that construct.  The corresponding phenomena in an origin-centric analysis would be to include several points that are actually coincident, thereby placing greater emphasis upon recovering the distance of this composite from all other points within the scatter.
 
Origin-centric methods allow the scatter to be visualized much like a ``Dyson Swarm'' for a finite universe of N or fewer points. For example, a q=3 dimensional recovered scatter might be primarily viewed as a 2-dimensional small-scale conformal mapping where radial distance information is disregarded. Alternatively, the size of each of the N plotting symbols displayed on this sort of ``sky view" map could represent some monotone function of its recovered radius.

Nonnegative Matrix Factorization (NMF) techniques rely upon locating the origin so that all N initial points have p strictly nonnegative coordinates.  In other words, all initial points must lie on or within the boundaries of the ``first'' Cartesian $2^p$-ant of p-dimensional Euclidean space. Similarly, the q dimensions to be recovered using NMF are to be represented as vectors emanating from the origin that have strictly nonnegative direction cosines because they point exclusively into this same first Cartesian $2^p$-ant.

\section{Appendix: Affine Invariance and Commutativity}
\label{sec: affine invarcomm}

Linear algebra is typically used to manipulate data structures, consisting of vectors and matrices of real numbers, in multivariate statistical analyses.  Insights into the analytical geometry of such manipulations is provided by visualizing the data as consisting of a scatter of N points within a p-dimensional Euclidean space.

A fundamental property of Euclidean space is that distance is computed from coordinates relative to orthogonal axes using the Pythagorean theorem.

A N $\times$ p matrix of real numbers, $\vec Y$, is said to represent a nonsingular affine transformation of an initial N $\times$ p matrix of real numbers, $\vec X$, if and only if there exists a 1 $\times$ p ``translation" vector $\vec {a'}$ and a nonsingular p $\times$ p matrix $\vec B$ of real values such that
\begin{equation}
\vec Y = \vec {1a'} + \vec {XB}. \label{eqn: afftran}
\end{equation}
In particular, if any row of $\vec X$ consists of all zeros, the corresponding row of $\vec Y$ will be $\vec {a'}$. The notation of equation (\ref{eqn: afftran}) differs slightly from that of Obenchain(1971) simply because the $\vec X$ and $\vec Y$ matrices have been transposed here.

A statistic $S(Y)$, consisting of a scalar, vector or matrix of real numbers, will be said to be affine invariant if and only if its numerical value when computed after any nonsingular affine transformation has been applied to a dataset is identical to its value before such transformation:
\begin{equation}
S(\vec Y) = S(\vec X) \mbox{  ...for every translation $\vec a$ and nonsingular $\vec B$} \label{eqn: affinvar}
\end{equation}

The $\vec H$ matrix of (\ref{eqn: svdcenter}) is not uniquely determined. For example, any of the r columns of $\vec H$ (each N $\times$ 1) can always be multiplied by $-1$ if the corresponding row of $\vec {G'}$ (1 $\times$ p) is also multiplied by $-1$.  On the other hand, $\vec {HH'}$ is an idempotent and symmetric (N $\times$ N) matrix that is uniquely determined and corresponds to the orthogonal projection matrix (linear operator) that characterizes the column space of the $( \vec I - \vec {1\gamma'} ) \vec X$ matrix; see Rao(1973), pages 46-48.  In fact, Obenchain(1971) showed that $\vec {HH'}$ and $\vec D^{(2)}$ are equivalent ``maximal affine invariant'' statistics which can be viewed as a standardized, canonical form for a given (N $\times$ p) data matrix of ``centered'' scalar values subject to nonsingular linear $\vec B$ transformations, (\ref{eqn: afftran}).

The $\vec {HH'}$ and $\vec D^{(2)}$ matrices are unique but distinct, and each can be computed from the other.  PCA focuses upon reproducing $( \vec I - \vec {1\gamma'} ) \vec {XX'} ( \vec I - \vec {\gamma 1'} )$ by decomposing it into ordered and additive components. PCA fails to reduce dimensionality when applied to $\vec {HH'}$ because all of its components are equally good or bad and, thus, cannot be uniquely ordered.

By focusing upon reproducing $\vec D^{(2)}$ instead of $\vec {HH'}$, origin-centric methods can reduce the dimensionality of configurations that PCA cannot.  There appear to be at least two separate parts to the price-one-has-to-pay to achieve affine invariance properties that ignore vagaries caused by analyst choice of variable scalings and/or confounding between variables.  First of all, recovered configurations will not be additive; the optimal solution of dimension (q+1) usually does not exactly ``contain'' the optimal solution of dimension q.  Secondly, minimizing $\vec {Norm}^2$ of equation (\ref{eqn: sqdifnorm}) is a problem of constrained optimization subject to multiple local minima.  Fortunately, modern numerical search software, such as the ``optimx'' R-package of Nash and Varadhan(2011), is now available to researchers.

\section{Appendix: Detail on Data Centering and Affine Median Vectors}
\label{sec: center}

A N $\times$ p matrix of real numbers, $\vec Y$, will be said to be a ``centered'' version of a N $\times$ p matrix of real numbers, $\vec X$, if and only if
\begin{equation}
\vec Y = ( \vec I - \vec 1\gamma' ) \vec X, \label{eqn: center}
\end{equation}
where the 1 $\times$ N row vector $\vec \gamma'$ is any generalized inverse of the N $\times$ 1 column vector of all ones, $\vec 1$.  In other words, any 1 $\times$ N row vector of  real values that sum to 1, $\vec \gamma'1 = 1$, is a valid choice for $\vec \gamma'$ in (\ref{eqn: center}).

The ``centering'' operation of equation (\ref{eqn: center}) frequently yields $\vec Y$ coordinates that differ from those of the initial $\vec X$ matrix.  In these cases, the location of the implicit origin, $\vec 0$, has been shifted via a simple ``translation'' of $\vec Y$ coordinates.

The unique Moore-Penrose inverse of $\vec 1$, usually denoted by $\vec 1^+$, is the row vector with each of its N entries equal to 1/N.  This choice for $\vec \gamma'$ corresponds to centering the $\vec X$ matrix at its overall mean vector, i.e. at its traditional centroid.

When $\vec \gamma_1$ and $\vec \gamma_2$ are two possibly different choices for the $\gamma$ vector of (\ref{eqn: center}), the following equality is easily verified by direct multiplication and algebraic simplification:
\begin{equation}
( \vec I - \vec 1\gamma'_2 )( \vec I - \vec 1\gamma'_1 ) = \vec I - \vec 1\gamma'_2 .
\label{eqn: idempot}
\end{equation}
When $\vec \gamma_1 = \vec \gamma_2$, equation (\ref{eqn: idempot}) shows that all valid choices for the $\gamma$ vector of (\ref{eqn: center}) yield a centering matrix, $\vec I - \vec 1\gamma'$ , that is idempotent and, thus, corresponds to a geometrical projection in N space that generally is oblique.  In fact, the choice $\vec \gamma'_1 = \vec \gamma'_2 = \vec 1^+$ is the only choice that makes the $\vec I - \vec {11^+}$ matrix symmetric as well as idempotent, thus corresponding to a strictly orthogonal projection in N space.

Perhaps, an even more interesting result follows from equation (\ref{eqn: idempot}) when $\vec \gamma_1$ and $\vec \gamma_2$ are distinct generalized inverses.  In these cases, note that any valid, final choice of centering, as in equation (\ref{eqn: center}), will simply ``wipe-out'' and ``replace'' any and all previously applied choices of centering of form (\ref{eqn: center}).  The corresponding implication for (\ref{eqn: svdcenter}) is that changing one's choice of $\vec \gamma$ vector typically changes the $\vec H$ matrix so that it becomes orthogonal to the new $\vec \gamma$. 

To center an $\vec X$ matrix at the point corresponding to its i-th row, the i-th element of $\gamma$ would then be a 1 and all other elements would be null.

A multivariate median vector can be defined, as in Obenchain(1972a), using ``convex hull'' concepts to assure that the $\vec \gamma$ vector is an invariant function under all strictly nonsingular affine transformations of a given $\vec X$ scatter, i.e. where no systematic reductions in dimensionality are being enforced.  In direct analogy with the concept of a univariate median, successive ``exterior'' convex hulls (which are preserved under nonsingular affine transformations) of the scatter initially spanning p-dimensional Euclidean space would be determined and successively ``peeled away.''  Whenever two or more points on an exterior hull are coincident, only one such point is set aside (assigned a $\gamma$ weight of zero) at each stage of such peeling.  The final convex hull remaining at the end of this sequence will contain neither any coincident points nor any strictly interior points.  The affine invariant $\vec \gamma$ vector, which defines the corresponding affine commutative median vector, $\vec {\gamma'X} = \vec {-a'}$ of equation (\ref{eqn: afftran}), gives equal, positive weight to each of the unique points on the innermost hull and zero weight to all other points.

\section{Appendix: Restrictions to Distinct Points, possibly Weighted.}
\label{sec: coincident}

In this penultimate appendix, we comment on the role of mutually coincident points in determining the dimensionality of a given scatter.

Since coincident points clearly cannot increase dimensionality and always remain coincident in the affine reduction formulation of (\ref{eqn: rqaffrmd}), it makes good sense computationally to limit attention to scatters of only distinct, non-coincident points.  Since no reasonable method will fail to exactly reproduce the zero diagonal elements of the $\vec D^{(2)}$ matrix of (\ref{eqn: sqdistmtx}), minimizing the squared norm of equation (\ref{eqn: sqdifnorm}) rightly focuses on minimizing the sum of the $N(N-1)/2$ terms with $1 \leq i < j \leq N$, where N now denotes the total number of distinct points.  On the other hand, individual terms in this summation could certainly be differently ``weighted'' to account for the presence or absence of additional points coincident with either the i-th or the j-th distinct point.

Limiting attention to unique points and affine reduction methods also assures, when many variables are analyzed for a fixed number of points, that the maximum possible value of r will indeed equal $(N-1) \leq p$ in equation (\ref{eqn: svdcenter}).  For an arbitrary configuration of N unique points, one then knows in advance that $\vec {H'H}$ will ultimately become the $(N-1) \times (N-1)$ identity matrix as the number of columns of $\vec X$ is arbitrarily increased.  Furthermore, in this same limit, the corresponding $\vec {HH'}$ matrix will also approach the $\vec {I - \gamma\gamma^+}$ matrix, which is the N $\times$ N orthogonal projection matrix for the vector space orthogonal to $\gamma$.

\section{Appendix: Affine Invariance and the Restriction to p less than or equal to $N-1$.}
\label{sec: plarge}  

When $\vec {\gamma'} = \vec {1^+}$, the large-p asymptotic H-configuration is that of N points equally spaced in $N-1$ dimensions, which corresponds to a sparse, high-dimensional Dyson sphere with no point coincident with the origin.  Specifically, the squared distance between any 2 of these N points will be 2, and the squared distance from each to the origin will be $d^2_0$ = $(N-1)/N$.

When all of the elements of $\vec \gamma$ except one are zeros, the asymptotic configuration will depict that single point as being coincident with the origin while the remaining $N-1$ points are again equally spaced.  While the squared distance between any two of these $N-1$ points will still be 2, the squared distance from each of them to the origin point will now be $d^2_0$ = 1. The first $N-1$ rows of the asymptotic $\vec H$ matrix may then be any $(N-1) \times (N-1)$ orthogonal matrix (such as some permutation of the rows and columns of the $\vec I$ matrix), while the final row is null.  This configuration again spans $N-1$ dimensions and depicts another sparse, high-dimensional Dyson sphere as well as its origin point.     

\section{Acknowledgments}
\label{sec: acknow}

The author received extensive feedback from reviewers of his 1972 submission on this topic to {\em Biometrika} as well as his 1973 submission to {\em Journal of Multivariate Analysis}.  The current manuscript is brief and completely revised in organization and presention style.  The basic concepts outlined here are unquestionably sound, and our objective of incorporating affine invariance and commutivity properties into algorithms for reduction of multivariable dimensionality is not ``impossible.''  Thanks to modern statistical computing systems for graphical visualization and numerical search, our innovative origin-centric approach can be of much more ``practical interest'' today than it was 40+ years ago.

\section{References}
\label{sec: refs}

Dyson, FJ. Search for Artificial Stellar Sources of Infra-Red Radiation. {\em Science}, 1960, 131, 1667-1668.

Gabriel, KR. The Biplot Graphic Display of Matrices with Application to Principal
Component Analysis. {\em Biometrika}, 1971, 58(3), 453-467.

Gabriel, KR. Analysis of Meteorological Data by Means of Canonical Decomposition
and Biplots. {\em Journal of Applied Meteorology}, 1972, 11, 1071-1077.

Glaeser, EL. Researcher incentives and empirical methods. {\em Harvard Institute of Economic Research, discussion paper \#2122}. http://scholar.harvard.edu/glaeser 2006.

Gower, JC. Some distance properties of latent root and vector methods used in multivariate analysis. {\em Biometrika}, 1966, 53, 325-338.

Gower, JC. Adding a point to vector diagrams in multivariate analysis. {\em Biometrika}, 1968, 55, 582-585.

Gower, JC. Euclidean distance geometry, {\em Math. Sci.}, 1982, 1, 1-14.

Gower, JC. Properties of Euclidean and Non-Euclidean Distance Matrices {\em Linear Algebra and its Applications}, 1985, 67, 81-97.

Lee, DD and Seung, HS. Learning the parts of objects by non-negative matrix factorization. {\em Nature}, 1999, 401(6755), 788-791.

Longley, JW.  An appraisal of least squares programs for the electronic computer from the point of view of the user.  {\em Journal of the American Statistical Association}, 1967,  62, 819-841.

Nash, JC and Varadhan, R. Unifying Optimization Algorithms to Aid Software
System Users: optimx for R.  {\em Journal of Statistical Software}, 2011, 43(9), 1-14.

Mahalanobis, PC. On the generalized distance in statistics. {\em Proc. Nat. Inst. Sci. India, Part A.} 1935, 12, 49-55.

Obenchain, RL. Multivariate procedures invariant under linear transformations. {\em Annals of Mathematical Statistics}, 1971, 42, 1569-1578.

Obenchain, RL. Correction Note: Multivariate procedures invariant under linear transformations. {\em Annals of Mathematical Statistics}, 1972a, 43, 1742-1743.

Obenchain, RL. Regression optimality of principal components.  {\em Annals of Mathematical Statistics}, 1972b, 43, 1317-1319.

Obenchain, RL. Affine Multidimensional Scaling. Presentation at {\em Third International Symposium on Multivariate Analysis}, Dayton, OH, 1972c.

Obenchain, RL. The geometry of linear reparameterizations in regression. {\em Contributions to Statistics: Essays in Honor of Norman L. Johnson}, ed. P. K. Sen, Amsterdam: North-Holland 1983, 347-365.

Obenchain, RL. ICE Preference Maps: Nonlinear Generalizations of Net Benefit and Acceptability.  {\em Health Services and Outcomes Research Methodology}, SpringerLink (open access), 2008, 8, 31-56. (two-dimensional, origin-centric maps of economic preferences.)

Okamoto, M. Optimality of principal components. {\em Multivariate Analysis, 2} P. R. Krishnaiah, ed. New York: Academic Press, 1969, 673-685.

Rao, CR.  Linear Statistical Inference and Its Applications, 2nd Edition.  New York: John Wiley and Sons, 1973.

Schoenberg, IJ. Remarks to Maurice Freehet's article ``Sur la definition axiomatique d'une classe d'espaces vectoriels distancies applicables vectoriellement sur l'espace de Hilbert'', {\em Ann. Math.} 1935, 36, 724-732.

Torgerson, WS. {\em Theory and methods of scaling.} New York: Wiley, 1958.

Young, G. and Householder, AS. Discussion of a set of points in terms of their
mutual distances. {\em Psychometrika}, 1938, 3, 19-22.

\end{document}